# Pitfalls and prospects of optical spectroscopy to characterize perovskite-transport layer interfaces

Eline M. Hutter,[1] Thomas Kirchartz,[3,4] Bruno Ehrler,[1] David Cahen,[5,6] Elizabeth von Hauff[1,2]

1. Center for Nanophotonics, AMOLF, Science Park 104, 1098 XG Amsterdam, the Netherlands
2. Department of Physics & Astronomy, VU Amsterdam, De Boelelaan 1081, 1081 HV Amsterdam, the Netherlands
3. IEK-5 Photovoltaik, Forschungszentrum Jülich, 52425 Jülich, Germany
4. Faculty of Engineering and CENIDE, University of Duisburg-Essen, Carl-Benz-Str. 199, 47057 Duisburg, Germany
5. Department of Materials and Interfaces, Weizmann Institute of Science, Rehovoth 76100, Israel
6. Dept. of Chemistry and Bar-Ilan Inst. of Nanotechnol. & Adv. Mater., Bar-Ilan Univ., Ramat Gan, Israel

**Introduction**

Over the last decade, metal-halide perovskite photovoltaics (PV) have demonstrated an unprecedented increase in power conversion efficiencies, from ~ 4%[1] in 2009 to ~ 25% in 2019.[2] This impressive learning curve is partially due to the unique material properties of perovskite absorber layers, such as the low density of defect states in the bandgap[3]. This low defect density is particularly remarkable given that the films used for photovoltaics are polycrystalline layers processed from solution near room temperature. The low defect density yields sharp absorption onsets and relatively high emission quantum yields ($QY$)[4,5]. Furthermore, tuning of the optical bandgap via varying the precursor composition allows solar cell structures (tandem) that harvest more of the energy in the solar spectrum. On the other hand, progress must also be attributed to the enormous global efforts by researchers with a variety of expertise, towards the fabrication, characterization, and engineering of high-performance energy-conversion and light-emitting devices.

The commercial feasibility of a PV technology can be qualitatively assessed based on the interplay between efficiency, cost, and lifetime. Perovskite solar cells score high on the first two points, so that improving material and device stability remains one of the largest challenges for the field. The choice of *transport layers*, through which photo-generated electrons and holes are transported to the electrodes, is a crucial factor for device performance[6,7] and stability.[8] Analogous to the development history of other PV technologies, this critical step in device engineering has motivated the perovskite PV community to search for ideal transport layer materials to prevent performance losses during charge extraction. The concept of *selective contacts* is well-known in PV,[8–11] in particular for the doped silicon transport layers in high performance silicon PV.[12] In contrast, the development of selective contacts in thin film PV has been more challenging.[13] This is because the requirements for a transport layer to be considered *selective* are strict; selective transport layers must be ohmic for the selected carriers, and prevent electrical losses, caused *e.g.* by a decrease in carrier mobility at the contact,[14] interfacial non-radiative recombination due to trapping, and minority carrier recombination. In addition to the requirement of selectivity, transport layers also have to be designed to minimize optical losses such as parasitic absorption, need to be minimized.[15] All of these losses will ultimately reduce the power conversion efficiency of the solar cell. Sometimes the term "extraction layer" is used in connection with transport materials that facilitate charge collection from the active layer. However, it is important to note that selective transport layers are *passive*, *i.e.* they simply form a loss-less contact between the semiconductor absorber layer and the metal contact for one of the charges (electron or hole), while the other charge is blocked by a sufficiently large barrier ($\Delta E \gg kT$). In other words, an ideal selective contact performs the simple, yet elusive, task of prohibiting electrical (non-radiative recombination) losses during carrier extraction.

In the case of thin-film PV such as perovskite PV, where the interfaces between the absorber and transport layers are formed by two different materials (heterojunction), a physico-chemical understanding of the interfacial properties is a prerequisite towards the rational design of efficient and stable solar cells. Therefore, engineering optimized architectures requires understanding of the energy level alignment and electric field distribution within the solar cell, as well as of the recombination channels in the absorber layer and at the device interfaces.

In this perspective, we specifically address the question of how to screen and characterize the quality of the interfaces between transport layers and perovskite absorbers. *In-situ* characterization tools are especially suitable for studying the optoelectronic properties of solar cell interfaces under conditions that are relevant for real applications, such as variations in absorber composition, different environmental stress factors, and ageing during device operation. We highlight the advantages and challenges in using photoluminescence (PL) spectroscopy, also in conjunction with other optical spectroscopies, to specifically target device interfaces for the screening of new transport layers. Selective transport layers should ideally not induce additional loss channels, and hence not affect the photoluminescence quantum yield ($Q_e^{PL}$) nor lead to a reduction in the energy of the excited state. As a result, the $Q_e^{PL}$ and spectral shape of the emission from the perovskite active layer should remain unchanged when interfaced with the transport layer. This means that, in principle, a steady-state, or continuous wave (cw), PL measurement may be sufficient to screen the passivation quality of new transport layers. However, using both steady-state and time-resolved (tr) spectroscopy, and combining these with electrical measurements is more suitable to study the contacts' selectivity and to derive quantitative information about the opto-electronic processes in the sample, including carrier densities, lifetimes, and charge-transfer rates at material interfaces. Such an approach is needed not only for understanding, but also for guiding material and device fabrication protocols.

Here we highlight the power of complementary studies that use both cw-PL and tr-PL to understand non-radiative losses, and additional transient spectroscopies for characterizing the potential for loss-less carrier extraction at the solar cell interfaces.[16] Based on our discussion we make recommendations on how to extrapolate results from optical measurements to assess the quality of a transport layer, and its impact on solar cell efficiency.

**Assessing interfacial quality between the perovskite and transport layer: PL quenching as a signature of non-radiative losses**

Following the model originally proposed by Shockley and Queisser,[17] i.e the SQ model, an ideal solar cell has a steady-state photoluminescence emission quantum efficiency ($Q_e^{PL}$) of unity. The SQ model assumes that each absorbed photon produces one electron-hole pair that rapidly thermalizes to the band edges. At open circuit voltage ($V_{oc}$) no net photocurrent flows and the absorption of radiation is balanced by the emission of radiation at the bandgap energy; thus, in this limit, all photo-generated charge recombines radiatively. In real semiconductors, the chemical potential of the photo-generated electrons and holes in this quasi-equilibrium scenario can be represented by the splitting of the quasi-Fermi levels of the electrons and holes. This quasi-Fermi level splitting ($QFLS$) is directly related to the density of photo-generated electrons and holes,[18,19] and sets an upper limit for the $V_{oc}$ of the solar cell.

If the quasi-Fermi levels are flat across the absorber layer, then the luminescence efficiency depends on applied voltage, but not on the source of excitation. Thus, the external photoluminescence and electroluminescence quantum efficiencies will be the same ($Q_e^{PL} = Q_e^{EL}$). In this scenario, any loss in the solar cell $V_{oc}$ is quantitatively correlated to losses in the external luminescence quantum efficiency $Q_e^{PL}$ via[20]

$$V_{oc} = V_{oc}^{rad} + \frac{kT}{q}\ln(Q_e^{PL}) \qquad (1)$$

meaning that each decade of loss in $Q_e^{\mathrm{PL}}$ relates to a voltage loss of about 60 meV. Here, *kT/q* is the thermal voltage (25.8 meV at 298 K, room temperature, RT) and $V_{oc}^{\mathrm{rad}}$ is the open-circuit voltage in the radiative limit (*i.e.* if all charges undergo radiative recombination).[15]

Further, in the limit of flat quasi-Fermi levels across the absorber, we can quantify the relation between the external luminescence quantum efficiency $Q_e^{\mathrm{PL}}$ and the internal luminescence quantum efficiency $Q_i^{\mathrm{lum}}$. The internal luminescence quantum efficiency is the ratio of the radiative and total recombination rates ($Q_i^{\mathrm{lum}} = R_\mathrm{R}/R_\mathrm{tot}$) and is related to $Q_e^{\mathrm{PL}}$ via the out-coupling efficiency $p_\mathrm{e}$ and the probability $p_\mathrm{r}$ of photon reabsorption by the absorber, which can lead to photon recycling:[15]

$$Q_e^{\mathrm{PL}} = \frac{p_e Q_i^{\mathrm{lum}}}{1 - p_r Q_i^{\mathrm{lum}}}. \qquad (2)$$

Thus, to achieve the ideal $V_{\mathrm{oc}}$ predicted by the radiative limit, every photon emitted by the absorber layer has to be either out-coupled or reabsorbed by the absorber layer itself.[21,22]

In addition to non-radiative recombination, substantial losses can occur from the parasitic absorption of (emitted) photons by the transport materials or metal electrodes. In real solar cells, quenching of the cw-PL resulting in $Q_e^{PL} < 1$ is associated with non-ideal, *i.e.* non-radiative losses, and therefore a reduction in $V_{\mathrm{oc}}$. In the literature, many studies demonstrate relatively good $Q_e^{PL}$ values for perovskite layers deposited onto insulating substrates, such as glass. These layers are often passivated either by controlled exposure to oxygen or by covering them with passivation layers, such as poly(methyl-methacrylate) (PMMA) or n-trioctylphosphine oxide (TOPO).[23,24] In contrast, there are few cw-PL studies showing high $Q_e^{PL}$ from perovskites interfaced with suitable *conductive* transport layers.[25,26] While there is little question that high $Q_e^{\mathrm{PL}}$ is indicative of high material quality in perovskite thin films on glass, the correct interpretation of a decrease in the cw-PL intensity for a perovskite layer interfaced with a conductive charge-transport layer, is less obvious.

In fact, in the case of perovskite-transport layer interfaces, PL quenching has led to contradicting interpretations. Some reports interpret PL quenching as a signature of efficient charge transfer,[27] while other reports correlate reduced PL intensity to non-radiative losses, such as surface recombination.[25,28] Part of this confusion may be linked to the fact that in the field of organic photovoltaics, PL quenching has long been interpreted as a sign of efficient charge separation at the molecular donor-accept interface. However, also in these systems quenching of the singlet emission from the donor molecule corresponds to a loss in carrier energy because it is associated with the formation of the localized, weakly emissive charge transfer state at the donor-acceptor interface.[29,30,31] This means that in addition to loss in carrier energy, there is also a loss in emission from the device. Both effects ultimately limit the solar cell $V_{oc}$.

Clearly, in the context of the radiative limit, *device interfaces should not quench the PL* of the absorber layer, so as to allow high open circuit voltages from devices. Further, charge transfer and extraction cannot be probed under steady-state open-circuit conditions, under which the net current flow in the device is zero. Instead, charge extraction can either be probed using PL spectroscopy away from $V_{oc}$,[32,33,34] or, as we discuss later, by using other transient techniques. This means that cw-PL performed at open-circuit conditions can only yield insight into how a given transport layer influences $Q_e^{PL}$ of the perovskite absorber. Insulating, passivating layers[24] may yield high $Q_e^{PL}$ but will not allow charge extraction. Only the combination of both high $Q_e^{PL}$ and low series resistance of the device is a sufficient condition for establishing that a given transport layer forms a suitable selective contact with the perovskite absorber. We note that fill factors around 80% in combination with negligible PL quenching have already been reported for a layer stack, using poly(triarylamine) (PTAA) as the hole transport material and PCBM as the electron transport material,[26] or with a passivating monolayer, self-assembled on oxide-based transport layers.[35] Thus, a very good compromise between efficient charge extraction and good surface passivation is fundamentally possible.

The use of cw and tr-PL to study and distinguish between ideal and non-ideal recombination processes is described in more detail below. In general, the excitation intensities and conditions used for transient and continuous excitation methods are quite different, and so are the subsequent carrier generation and recombination rates. Therefore, comparing the results, and subsequently drawing conclusions about processes occurring under standard solar cell working conditions requires care.[16]

**Predicting $V_{oc}$ from time-resolved photoluminescence**

While cw-PL is a fast way to screen non-radiative losses, tr-PL can yield insights into the rates of these processes.[36] These rates can then be used to predict the upper limit for the device $V_{OC}$ as a function of perovskite composition or its contact with a specific charge transport layer, as detailed below.[37–40] First, it is important to understand that the excitation density has a significant effect on the recombination lifetime, and for this reason transient measurements are commonly performed as function of excitation intensity. In transient measurements, the time-dependent values of the photo-generated electron density $\Delta n$ (hole density $\Delta p$) take the general form

$$\frac{d\Delta n(t)}{dt} = G - R_R - R_{NR} + G_{rec} = G - k_R \Delta n(t) \Delta p(t) - k_{NR} \Delta n(t) + p_r k_R \Delta n(t) \Delta p(t) \quad (3)$$

where $G$, $R_R$ and $R_{NR}$ represent the rates of charge carrier generation, radiative recombination and non-radiative recombination, respectively, as schematically depicted in Figure 1. Recycling of emitted photons leads to an additional generation term $G_{rec}$ (= $p_r R_R$). Equation 3 assumes an intrinsic semiconductor. This is a fair assumption for halide perovskites, in which the background doping levels are typically much smaller (<$10^{12}$ cm$^{-3}$)[41] than the excitation density at 1 sun ($10^{15}$ cm$^{-3}$).[39] However, the model will be more complicated for highly doped semiconductors. All rates are dependent on the excitation density ($\Delta n$), as shown in Figure 2a, but the corresponding **rate constants** $k_R$ and $k_{NR}$ are not time- or excitation density-dependent. The rate constants can be determined by cw or tr-PL as described below.[36,38,39]

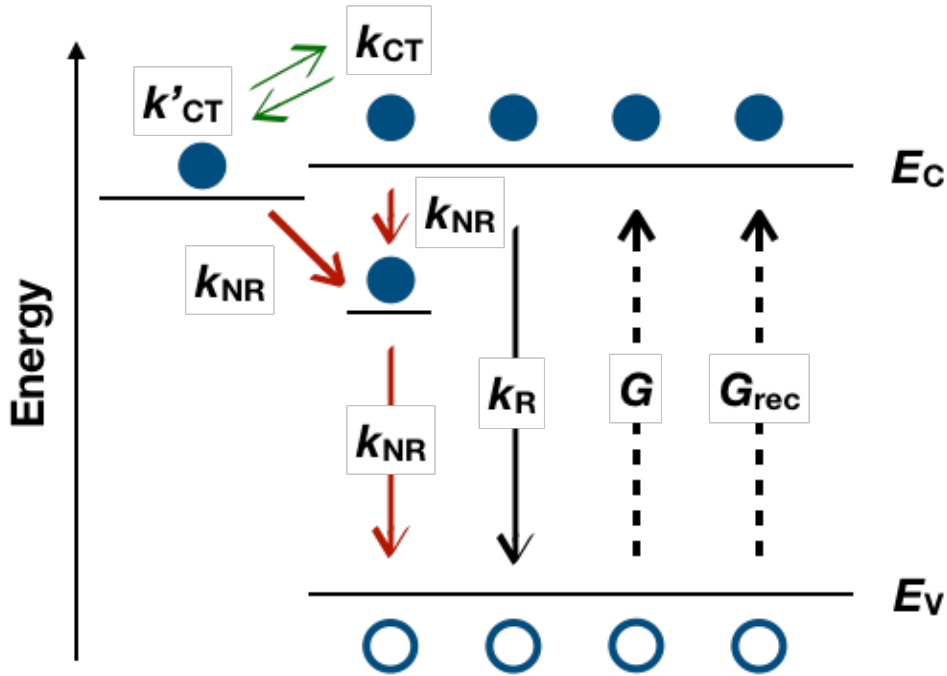

**Figure 1: Optical excitation of the absorber leads to generation ($G$) of photocarriers. Carriers ideally recombine radiatively ($k_R$) via band-to-band recombination. Photon recycling results in further generation of carriers ($G_{rec}$). At the interface between the perovskite and transport layer, charge transfer ($k_{CT}$) processes that are associated with corresponding back transfer of charge ($k'_{CT}$) do not reduce the cw-PL intensity, but will affect $\Delta n(t)$ and may thus be observed in the tr-PL spectrum. Non-radiative losses ($k_{NR}$), such as defect-assisted recombination due to traps in the bulk active layer or at the interface between the active layer and the transport layer will reduce the PLQY.**

At excitation densities well above the trap density, the non-radiative losses in $\Delta n$ are relatively small ($R_R > R_{NR}$) so that $\Delta n \approx \Delta p$. Hence, if radiative recombination dominates (high carrier density), then $d\Delta n(t)/dt \propto \Delta n^2$, while in the non-radiative regime, $d\Delta n(t)/dt \propto \Delta n$ (low carrier density), as shown in Figure 2 a).

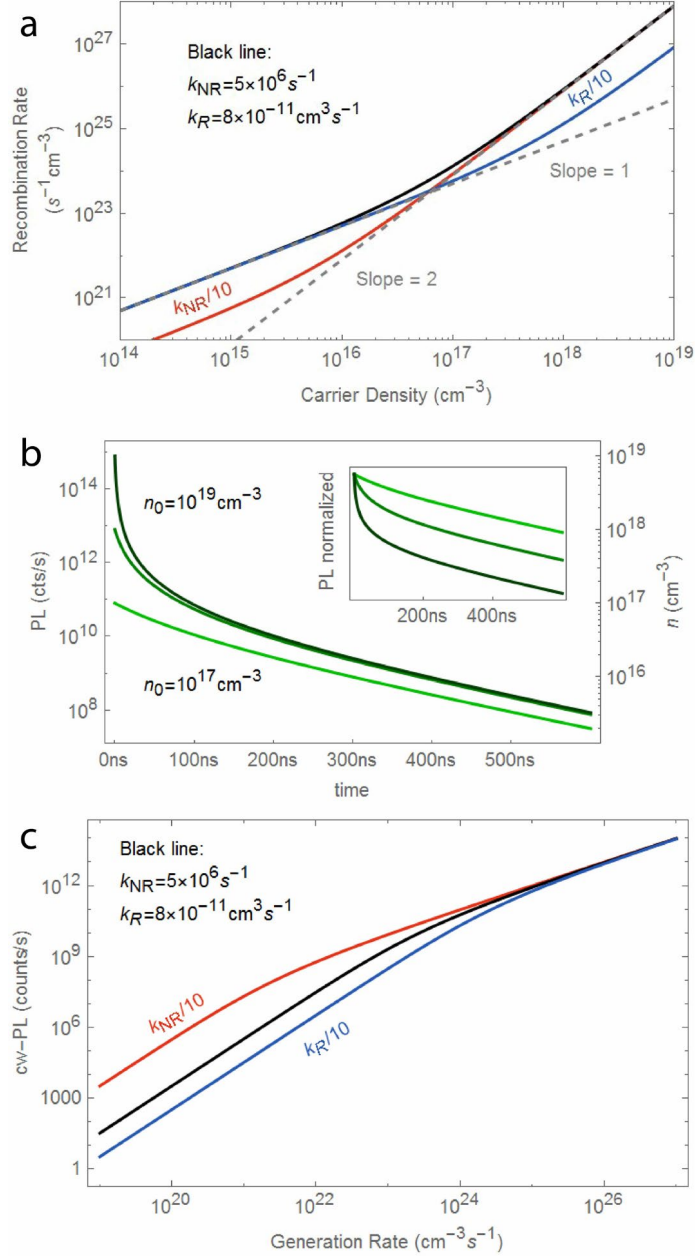

**Figure 2: Simulated photoluminescence following equation 3 and 4. a) Recombination rate ($k_R n^2 + k_{NR} n$) for different carrier densities within the perovskite layer. The slope goes from 1 in the low-injection (non-radiative) regime to 2 in the high-injection (radiative) regime. Values for the black line are taken from Ref. 23. b) tr-PL curves for different initial carrier densities. The quadratic regime is only visible at high density. c) cw-PL as simulated by $k_R n^2$ for different steady-state generation rates.**

In a tr-PL measurement the initial carrier density decays over time and the measurement thus samples over a range of carrier densities, as visualized in Figure 2b, and the rate constant can be obtained from fitting the decays with Equation 3. It should be noted that non-radiative recombination of charges reduces $\Delta n$(t) and consequently lowers the rate of radiative recombination ($R_R$). Thus, a long PL lifetime does not necessarily indicate that all photo-generated charges are long-lived, as this may also be observed if the majority of charges is trapped.[42,43] Therefore, multiple excitation densities must be measured to get insight in the dominant recombination pathways. The $Q_e^{PL}$ gives additional insight in whether a long PL lifetime is associated with non-radiative losses.

The main difference between cw-PL and tr-PL is that the carrier density in the steady-state measurement is fixed by the competition of generation and recombination rates. In quasi-equilibrium:

$$\Delta n = \frac{k_{NR}}{2\,k_R}\left(-1 + \sqrt{1 + \frac{4Gk_R}{k_{NR}^2}}\right) \qquad (4)$$

This means that in cw-PL, the intensity versus fluence (directly related to $G$) is quadratic in the *non-radiative* regime and linear in the radiative regime, see Figure 2c. Thus, the fluence-dependence of cw-PL can in principle also be used to extract the values for $k_R$ and $k_{NR}$. However, as visualized in Figure 2c, this often requires a very large range of excitation densities. It is typically easier to obtain the rates from tr-PL as this requires only a few different initial excitation densities to extract the recombination rates (*i.e.* from the variation in lifetime, recall Figure 2b). We note that in mixed-halide perovskites, where phase segregation can lead to an inhomogeneous energy landscape, photodoping occurs.[44] This additional doping can lead to first-order radiative recombination.

The rate constants $k_R$ and $k_{NR}$ obtained from fitting the tr-PL data can be used to determine $\Delta n$ and $\Delta p$ in steady-state (*i.e.* $V_{oc}$) conditions. Here, $d\Delta n(t)/dt=0$, and the continuous illumination results in $G$ being constant in time and thus, constant in values of $\Delta n$ and $\Delta p$ for a given illumination intensity. From here, the *QFLS* can be calculated, which gives an upper limit for $V_{oc}$. This approach therefore enables screening of various perovskite materials with transient excitation (and detection) techniques; this information can then be used to estimate the optimum $V_{oc}$ in the corresponding devices.[19,25]

**Gaining insight into non-radiative pathways from other time-resolved spectroscopy techniques**

To elucidate the nature of non-radiative recombination pathways at the perovskite-transport material interface, transient absorption spectroscopy (TAS)[37] or time-resolved conductivity (measured via microwave or THz absorption)[45] measurements are more suitable than tr-PL. While tr-PL offers insight into different radiative pathways and decreases in the PLQY due to non-radiative losses, it does not allow for distinguishing between different non-radiative pathways. Therefore, the combination of PL studies (both cw and tr) and other transient techniques is a powerful way to screen new materials for application as selective contacts in perovskite solar cells. However, we note that different transient techniques are typically performed at different excitation densities, which may make it difficult to extrapolate the results to the operational conditions of the solar cell.[16] For example, excitation densities in TAS or THz spectroscopy measurements on perovskites are typically >$10^{17}$ cm$^{-3}$, which is several orders of magnitude larger than the ~$10^{15}$ cm$^{-3}$ obtained by AM1.5 illumination. Hence, the recombination processes often observed with TAS or THz spectroscopy, such as third-order Auger recombination (~$\Delta n^3$, ignored in Eq. 3), are not necessarily dominating in the corresponding solar cell.

Time-resolved microwave conductivity (TRMC) and tr-PL, on the other hand, are useful techniques to access the relevant, lower excitation regimes of $10^{14}$ to $10^{16}$ cm$^{-3}$. TRMC measurements determine the charge carrier mobility and the lifetime of *mobile* charges, and can thus be used to predict their diffusion in either the perovskite or the transport layer by measuring these layers individually.[38,46] The combination of TRMC and tr-PL is especially useful, since TRMC probes the recombination of all mobile charges, *i.e.* both radiative and non-radiative, while tr-PL selectively measures the radiative recombination lifetimes. Extensive analysis is however needed to differentiate the underlying physical processes corresponding to the observed kinetics. In a bilayer consisting of a perovskite and a transport layer for instance, a decreased PL lifetime could be due to charge transfer at the interface, but it can also be caused by fast trapping of charges by defects at the interface. The TRMC lifetimes distinguish between these processes, since efficient charge separation results in infinitely long lifetimes of mobile charges, whereas interfacial recombination rapidly quenches mobile charges resulting in sub-ns lifetimes.[45] Finally, with TAS measurements, the bleach of each layer that is populated with photo-generated charges can be probed as a function of time after excitation. This spectral resolution allows us to differentiate between charge dynamics in the perovskite and the transport layer. Hence, TAS can be used to discriminate between charge transfer, interfacial recombination and trapping either inside the perovskite layer or in the transport layer, provided that the excitation densities are low enough and the energetic landscape is known.[37] Including non-radiative loss processes in the charge-transport layers in $R_{NR}$ (Eq. 3) enables us to predict the $V_{oc}$ losses occurring both in the perovskite layer itself and at the different interfaces.

**Routes towards finding selective transport layers in perovskite solar cells**

In table 1 we summarize our discussion of cw-PL and transient techniques that can be applied for screening perovskite-transport layer structures, and discriminating between radiative and non-radiative processes. The combination of cw- and tr-PL is a powerful approach to study both the magnitude (*i.e.* intensity in cw-PL) and rate (*i.e.* lifetime in tr-PL) of radiative recombination, and will yield quantitative insight into non-radiative processes induced by interfacing the perovskite with a specific transport layer. Set-ups for cw-PL measurements are available in most device fabrication laboratories, and can be quickly applied to screen whether a transport layer material induces changes in the perovskite PL. Considering that any quenching of the cw-PL by a transport layer means a loss in $V_{oc}$, the intensity and spectrum of cw-PL already gives a good indication whether a transport layer introduces losses in the charge-carrier density or energy, respectively. However, the reduction in PL gives very limited insight in the loss channels and it is not always possible to measure enough excitation densities with cw-PL to obtain quantitative information. In principle, cw-PL measurements can also be performed in full devices, away from $V_{oc}$ conditions in order to identify losses occurring during carrier extraction.[32] Still, complete devices contain multiple interfaces, and isolating the influence of a single transport layer on performance losses is non-trivial.

With tr-PL, each interface can be assessed individually, by comparing the transients of perovskite layers and perovskite/transport layer bilayers. In addition, it is much easier to get information about the charge-carrier dynamics with tr-PL measurements than with cw-PL. However, the interpretation of tr-PL is not trivial, and multiple excitation densities and kinetic models are needed to understand the underlying recombination processes. Fitting the transients yields the rate constants for radiative and non-radiative recombination, which can be used to predict the upper limit of $V_{oc}$ for a certain perovskite/transport layer combination. Finally, more advanced (but less available) tr-spectroscopy techniques such as TAS, TRMC, and THz spectroscopy can yield further insights into non-radiative carrier dynamics. Care, however, must be taken due to the differences in the excitation densities used for each measurement.

| Technique | Reveals | Pros & Cons |
|---|---|---|
| cw-PL | <ul><li>Emission energy</li><li>PLQY</li><li>luminescence quenching</li><li>$k_R$ and $k_{NR}$ for a wide range of illumination conditions</li></ul> | + Simple<br>+ Spectral information<br>+ No full device needed<br>- Wide range of intensities needed to get rate constants/not easy to access quantitative information<br>- Loss channels not defined<br>- Limited information |
| TRPL | <ul><li>Lifetime quenching</li><li>Sensitive measurement of $k_R$ and $k_{NR}$ for layers without contacts</li><li>Information about surface recombination velocities for layers with contacts</li><li>Need to make sure density is suitable to measure both rates</li></ul> | + Relatively simple<br>+ No full device needed<br>- Interpretation not trivial<br>- May require complementary measurements<br>- (numerical); model is needed, especially for layers with contacts |
| TAS | <ul><li>Timescale for all processes</li><li>Transfer across junctions</li><li>Emission energy</li><li>ps to ns timescales</li></ul> | + Spectral information<br>+ Probes non-radiative decay<br>+ No full device needed<br>- Complex measurement<br>- High excitation density |
| TRMC | <ul><li>Lifetime and mobility of free charges</li><li>ns to microsecond timescale</li></ul> | + Distinguish between charge transfer and interfacial recombination<br>+ Low excitation density<br>+ No full device needed<br>- Limited availability<br>- Can only extract product of carrier density and mobility |
| THz spectroscopy | <ul><li>Lifetime and mobility of free charges in perovskite</li><li>Sub-ns timescales</li></ul> | + Distinguish between charge transfer and interfacial recombination<br>+ no full device needed |

| | | - Very high excitation density<br>- Limited availability<br>- Can only extract product of carrier density and mobility |
|---|---|---|

Currently, most state-of-the-art transport layers introduce non-radiative losses, as evidenced by the frequently-observed PL quenching in bilayers of perovskite/transport layer, which significantly reduces the $QFLS$ and thus the $V_{oc}$.[19,25] Another limitation of frequently used (organic) transport layers in perovskite solar cells is their low conductivity, which limits the charge transport to the electrodes, reducing the fill factor ($FF$) of the device.[14] Ideally, one would use a semiconducting transport layer and tune its conductivity via controlled doping, or else a very thin buffer layer, to avoid electrical losses. However, this is complicated by the fact that halide perovskites tend to react with metals and metal oxides[47,48] without suitable functionalization.[49] Therefore, to maximize both $FF$ and $V_{oc}$, future work should focus on optimizing the conductivity of the transport layer, while minimizing chemical reactivity and interfacial recombination.

## Acknowledgements

The work of E.M.H. and B.E. is part of the Dutch Research Council (NWO) and was performed at the research institute AMOLF. TK acknowledge the Helmholtz Association for funding via the PEROSEED project.